\documentclass[iop]{emulateapj}
\usepackage[flushleft]{threeparttable}

\shorttitle{Parsec-scale jets in LLAGN}
\shortauthors{Mezcua \& Prieto}

\begin{document}

\title{Evidence of parsec-scale jets in \\Low-luminosity Active Galactic Nuclei}

\author{M.~Mezcua\altaffilmark{1,2} and M.~A.~Prieto\altaffilmark{1,2 $\dagger$}}
\email{$\dagger$Invited guest at the Max-Planck-Institut f\"ur Extraterrestrische Physik, Garching, Germany}

\affil{$^{1}$Instituto de Astrof\'isica de Canarias (IAC), E-38200 La Laguna, Tenerife, Spain;  mmezcua@iac.es}
\affil{$^{2}$Universidad de La Laguna, Dept. Astrof\'isica, E-38206 La Laguna, Tenerife, Spain}


\begin{abstract}
The nuclear radio emission of low-luminosity active galactic nuclei (LLAGN) is often associated with unresolved cores. In this paper we show that most LLAGN present extended jet radio emission when observed with sufficient angular resolution and sensitivity. They are thus able to power, at least, parsec-scale radio jets. To increase the detection rate of jets in LLAGN, we analyze subarcsecond resolution data of three low-ionization nuclear emission regions. This yields the detection of extended jet-like radio structures in NGC\,1097 and NGC\,2911 and the first resolved parsec-scale jet of NGC\,4594 (Sombrero). The three sources belong to a sample of nearby LLAGN for which high-spatial-resolution spectral energy distribution of their core emission is available. This allows us to study their accretion rate and jet power ($Q_\mathrm{jet}$) without drawing on (most) of the ad hoc assumptions usually considered in large statistical surveys. We find that those LLAGN with large-scale radio jets ($>$100 pc) have $Q_\mathrm{jet} > 10^{42}$ erg s$^{-1}$, while the lowest $Q_\mathrm{jet}$ correspond to those LLAGN with parsec-scale ($\leq$100 pc) jets. The $Q_\mathrm{jet}$ is at least as large as the radiated bolometric luminosity for all LLAGN, in agreement with previous statistical studies. Our detection of parsec-scale jets in individual objects further shows that the kinematic jet contribution is equally important in large- or parsec-scale objects. We also find that the Eddington-scaled accretion rate is still highly sub-Eddington ($<10^{-4}$) when adding the $Q_\mathrm{jet}$ to the total emitted luminosity (radiated plus kinetic). This indicates that LLAGN are not only inefficient radiators but that they also accrete inefficiently or are very efficient advectors.

\end{abstract}


\section{Introduction}
\label{intro}
The presence of an accreting engine in the active galactic nuclei (AGN) with the lowest luminosities (LLAGN; $L_\mathrm{bol}\leq10^{42}$ erg s$^{-1}$) is a matter of debate (see review by \citealt{2008ARA&A..46..475H}).  
LLAGN are found at the center of nearly one-third of all local galaxies and encompass low-luminosity Seyferts, low-ionization nuclear emission regions (LINERs; \citealt{1980A&A....87..152H}), and transition nuclei (which have spectral properties intermediate between those of LINERs and HII regions).
LINERs are characterized by very strong low-ionization optical forbidden lines whose source of ionization has been debated for decades. The main scenarios include a low-luminosity version of bright AGN, shock heating (e.g., \citealt{1995ApJ...455..468D}), or stellar photoionization (e.g., \citealt{1985MNRAS.213..841T}; \citealt{1992ApJ...397L..79F}). The spectral energy distribution (SED) of LLAGN is very prominent in the radio band and lacks the canonical optical-UV ``big blue bump'' of AGN (indicative of the presence of an optically thick, geometrically thin, accretion disk). A combination of an optically thin advection-dominated accretion flow (ADAF; \citealt{1994ApJ...428L..13N}; see \citealt{2014arXiv1401.0586Y} for a review), an outer truncated disk (\citealt{1999ApJ...525L..89Q}), and a radio jet (\citealt{1995A&A...293..665F,1999A&A...342...49F}; \citealt{1996ApJ...464L..67F}) was proposed to model the SED of LLAGN (e.g., \citealt{1999ApJ...516..672H}; \citealt{2001ApJ...562L.133U}; \citealt{2006ApJ...643..652N,2012Sci...338.1445N,2014MNRAS.438.2804N}; \citealt{2010ApJS..187..135E}; \citealt{2011ApJ...726...87Y}; \citealt{2012JPhCS.372a2006F}; although see also \citealt{2013ApJ...777..164M} and references therein). However, the ADAF model alone is not able to satisfactorily reproduce the observed radio luminosities of LLAGN (e.g., \citealt{2003ApJ...588..175F}; \citealt{2005MNRAS.360..119D}a; \citealt{2005ApJ...621..130W}; \citealt{2007ApJ...669...96W}; \citealt{2008ApJ...681..905M}) and predicts an inverted radio spectral index ($\alpha\sim$0.4\footnote{The spectral index $\alpha$ is defined from $S_{\nu} \propto \nu^{\alpha}$, where \textit{S} is the flux density at frequency $\nu$.}; \citealt{1998ApJ...499..198Y}) much higher than the one typically observed in LLAGN spectra ($\alpha\sim -0.2$ to 0.2; e.g., \citealt{2000ApJ...542..197F}; \citealt{2001ApJ...559L..87N}; \citealt{2005MNRAS.363..692D}b; \citealt{2011AJ....142..167D}). Although this does not rule out the ADAF model, it argues for the necessity of a parsec-scale (pc-scale) radio jet to model the SEDs of these sources.

Despite the results showing that the SED of LLAGN is compatible with jet-dominated emission, observational evidence of radio jets in these sources still remains elusive. Several studies have aimed at detecting jet radio emission in LLAGN, finding that they are usually associated with compact radio sources where most of the radio emission is highly concentrated (e.g., \citealt{2000ApJ...542..197F}; \citealt{2000ApJS..129...93F}; \citealt{2001ApJ...562L.133U}; \citealt{2002A&A...385..425F}; \citealt{2004ApJ...603...42A}; \citealt{2005MNRAS.363..692D}b; \citealt{2005A&A...435..521N}; \citealt{2006A&A...451...71F}). Sub-parsec-scale radio emission was detected in 98\% LLAGN and AGN of the Palomar Sample observed with the VLBA\footnote{Very Long Baseline Array of the National Radio Astronomy Observatory (NRAO).} at 5 GHz (e.g., \citealt{2005A&A...435..521N}). The characteristic variability of LINERs, sometimes on timescales of months (e.g., \citealt{2002A&A...392...53N}; \citealt{2005ApJ...627..674A}; \citealt{2005ApJ...625..699M}; \citealt{2007MNRAS.377.1696M}; \citealt{2011A&A...527A.142G}), agrees with the confinement of the radio emission to compact cores. 
Most of these sources present a flat or slightly inverted radio spectrum and non-thermal brightness temperatures $T_\mathrm{B} > 10^{5}$ K footprint of a relativistic jet (\citealt{1979ApJ...232...34B}), indicating thus the presence of an AGN. However, only few of them show jet-like outflows or slightly resolved core emission at pc-scales (e.g., \citealt{2000ApJ...542..197F}; \citealt{2002A&A...392...53N}; \citealt{2005A&A...435..521N}; \citealt{2002A&A...385..425F}a; \citealt{2004A&A...418..429F}; \citealt{2007A&A...464..553K}; \citealt{2013ApJ...779....6H}), while larger kpc-scale radio jets in LLAGN is even scarcer (e.g., Cen A, \citealt{1992ApJ...395..444C}; M87, \citealt{2005AJ....130.1389L}; NGC\,1052, \citealt{1984ApJ...284..531W}; \citealt{2004A&A...420..467K}).
The question that arises is: do radio observations fail to detect radio jets in LLAGN due to insufficient angular resolution and sensitivity, or do most LLAGN show compact cores due to insufficient energy to launch a jet at parsec or larger scales?
It is clear that the answer must come from high-resolution and -sensitivity radio campaigns reaching sub-arcsecond resolutions, which are the only ones plausibly able to detect pc-scale jets in these sources. 

In this paper we analyze subarcsecond archival VLA\footnote{Very Large Array, now officially known as the Karl G. Jansky VLA, of the NRAO.} and VLBA data so far unpublished of a sample of nearby LLAGN studied in the IR at high spatial resolution by \cite{2010MNRAS.402..724P} and \cite{2010MNRAS.402..879R}, from now on PR2010. The results yield the first resolved radio structure of the pc-scale jet in NGC\,4594 (or Sombrero) and the detection of jet-like extended emission in NGC\,1097 and NGC\,2911.
We also compile and derive the jet power, bolometric luminosity, and Eddington accretion rates of the eight sources included in the sample of PR2010, which allows us to perform a study of the nuclear energetics of LLAGN devoid of most of the assumptions usually adopted in large statistical surveys. 

The objects and the radio data analyzed are presented in Section~\ref{data}, while the results obtained are shown in Section~\ref{results} and discussed in Section~\ref{discussion}. Final conclusions are summarized in Section~\ref{conclusions}.


\begin{table*}
\centering
\begin{threeparttable}
\caption{Targets and Radio Archival Data \label{table1}}
\begin{tabular}{lcccccc}
\tableline
\tableline
 Name	&  	  Array     &  Obs. date  &  Frequency   &   Resolution    	& Peak				&  rms   			\\
               &   		       &		     &    (GHz)	    &     (mas)         	& (mJy beam$^{-1}$) 	&   (mJy beam$^{-1}$)	        \\
    (1)       &   	   (2)        &     (3)       &   (4)              &        (5)		&     (6)                           &     (7)         	  \\
\hline
NGC\,1097 &  	 VLBA     &   2010 May  &       8.4        &    6.34 $\times$ 1.66      			    &		0.9           &  0.1	        \\
NGC\,2911 &   	 VLA-A     &  1986 Mar   &       4.9        & 0$\arcsec$.72 $\times$ 0$\arcsec$.35 &	70.1          &  0.2        \\
		  &   		       &  1986  Mar   &      14.9        & 0$\arcsec$.21 $\times$ 0$\arcsec$.11 &	 21.5          &  0.6         \\
NGC\,4594 &  	VLBA     &   2005 Jun  &       8.6          &    4.43 $\times$ 0.97      		      	    &	68.0             &  0.4        \\
		  &  		      &   2011 Mar   &       15.3        &    2.11 $\times$ 0.60      		            &	73.5            &    0.3        \\
		  &  		      &  2011 Mar   &       23.8        &    1.13 $\times$ 0.62      		           &	8.9              &   0.3        \\
\tableline
\end{tabular}
\begin{tablenotes}
      \small
      \item {\bf Notes.}~(1) Source name; (2) array configuration; (3) observing date; (4) observing frequency; (5) beam size; (6) peak flux density; (7) off-source rms noise.
         \end{tablenotes}
\end{threeparttable}
\end{table*}

\section{Sample and data analysis}
\label{data}
The three LINERs studied in this paper (NGC\,1097, NGC\,2911, and NGC\,4594; see Table~\ref{table1}) belong to the sample of the nearest and brightest LLAGN galaxies observed at high resolution in the IR with the Very Large Telescope (\citealt{2010MNRAS.402..724P}; \citealt{2010MNRAS.402..879R}). The proximity and brightness of these objects allow them to be studied at pc scales in the IR, optical/UV (e.g., \citealt{1995ApJ...440...91M}) and radio (e.g., \citealt{2010MNRAS.401.2599O}), yielding the highest spatial-resolution (subarcsecond) SEDs of their nuclei (\citealt{2012JPhCS.372a2006F}, in preparation). 

Archival VLBA data of NGC\,1097 and NGC\,4594 and VLA data of NGC\,2911 not previously published\footnote{The data for NGC\,4594 have also been reported by \cite{2013ApJ...779....6H} at the time of writing this paper.} were analyzed with the aim of resolving the large-scale radio emission and detecting a possible pc-scale jet in these sources. The A-array configuration of the VLA as well as frequencies $\geq$5 GHz were considered in order to achieve the subarcsecond resolution required to resolve the jet structure (see Table~\ref{table1}). The PR2010 sample also includes the LLAGN Cen~A, M87, NGC\,1052, NGC1386, and NGC\,3169. However, no unreported radio data at high-angular resolution was found in the archives for these sources.

The data were reduced following standard procedures with AIPS.\footnote{Astronomical Image Processing Software of NRAO.} The VLA data for NGC\,2911 were calibrated in amplitude using 3C\,286 as flux calibrator, while delay and gain solutions were derived from the phase calibrator 0851+202 and interpolated and applied to the target source.
The VLBA data for NGC\,1097 and NGC\,4594 were calibrated in amplitude using gain values and system temperatures. The uncertainties of this calibration are typically 5\% at the lower frequencies and 10\% at 22 GHz. The sources 3C\,279 (for NGC\,4594) and DA\,193 (for NGC\,1097) were used as fringe finders and bandpass calibrators. Delay, delay rate, and phase solutions were derived in the fringe-fitting using the phase calibrators J1239-1023 for NGC\,4594 and J0246-2935 for NGC\,1097, and were interpolated and applied to the target sources. 
Phase-referenced images of each target were produced by applying \texttt{CLEAN} deconvolution to the naturally weighted data. No self-calibration was applied. For NGC\,4594, the longest baselines were tapered to improve the detection of extended emission. The resulting restoring beams, peak flux densities, and rms noises are given in Table~\ref{table1}.

The total flux density, size, and position angle (P.A.) of the detected structures were measured from the lowest 3$\sigma$ contours using the AIPS tasks \texttt{TVSTAT} and \texttt{TVDIST}. The core components were, in addition, fitted by a two-dimensional elliptical Gaussian on the image plane using the AIPS task \texttt{JMFIT}. For NGC\,1097 and NGC\,4594, the core brightness temperature was derived from the size and flux density provided by the Gaussian fitting following the equation of \cite{1982ApJ...252..102C}.

\begin{table*}
\centering
\begin{threeparttable}
\caption{Results of the Radio Observations \label{table2}}
\begin{tabular}{lccccccc}
\tableline
\tableline
 Name	& Frequency   &  Core Flux	&  Total Flux   &     Size      	&  Size    &  P.A.    		&  $T_\mathrm{B}$  \\
               &  (GHz)	   &   (mJy)		&	(mJy)        &   (mas)     	&  (pc)     &  ($^{\circ}$)   	&	($\times 10^{6}$ K)  	\\
    (1)       &   (2)         &    (3)                 &   (4)             &        (5)		&    (6)     &   (7)     		&   (8)   \\
\hline
NGC\,1097 &  8.4         &  	1.0		&    1.1      &     8.7    		&   0.7      &  $-$23   & $>$2  \\
NGC\,2911 &  4.9         &		72.9		&   73.6        &    2$\arcsec$.6 	&  509   &   $-$54   & ... \\
		 &  14.9       &		21.0		&   25.6        &     0$\arcsec$.4  &  81     &  $-$90  &  ...  	 \\
NGC\,4594 &  8.6        &		75.4		 &   81.2        &      16.8   		&   0.7       &  $-$16  &  379  	\\
		  & 15.3        &		86.5		&   88.9         &     6.2  		&   0.3       &  $-$21    & 439  	 \\
		 & 23.8        &		16.9		&   36.7         &       14.4$^{a}$  	&   6       &   $-$31     & $>$40 	 \\
\tableline
\end{tabular}
\begin{tablenotes}
      \small
      \item {\bf Notes.}~(1) Source name; (2) observing frequency; (3) core flux density; (4) total flux density; (5) and (6) largest angular size as measured from the lowest 3$\sigma$ contour, in mas and pc respectively; (7) position angle of the extended jet structure; (8) brightness temperature. \\
$^a$ Size measured from component A to C, as labeled in Figure~\ref{ngc4594_23GHz}.
   \end{tablenotes}
\end{threeparttable}
\end{table*}

\section{Results}
\label{results}
\begin{figure*}
 \includegraphics[scale=0.38]{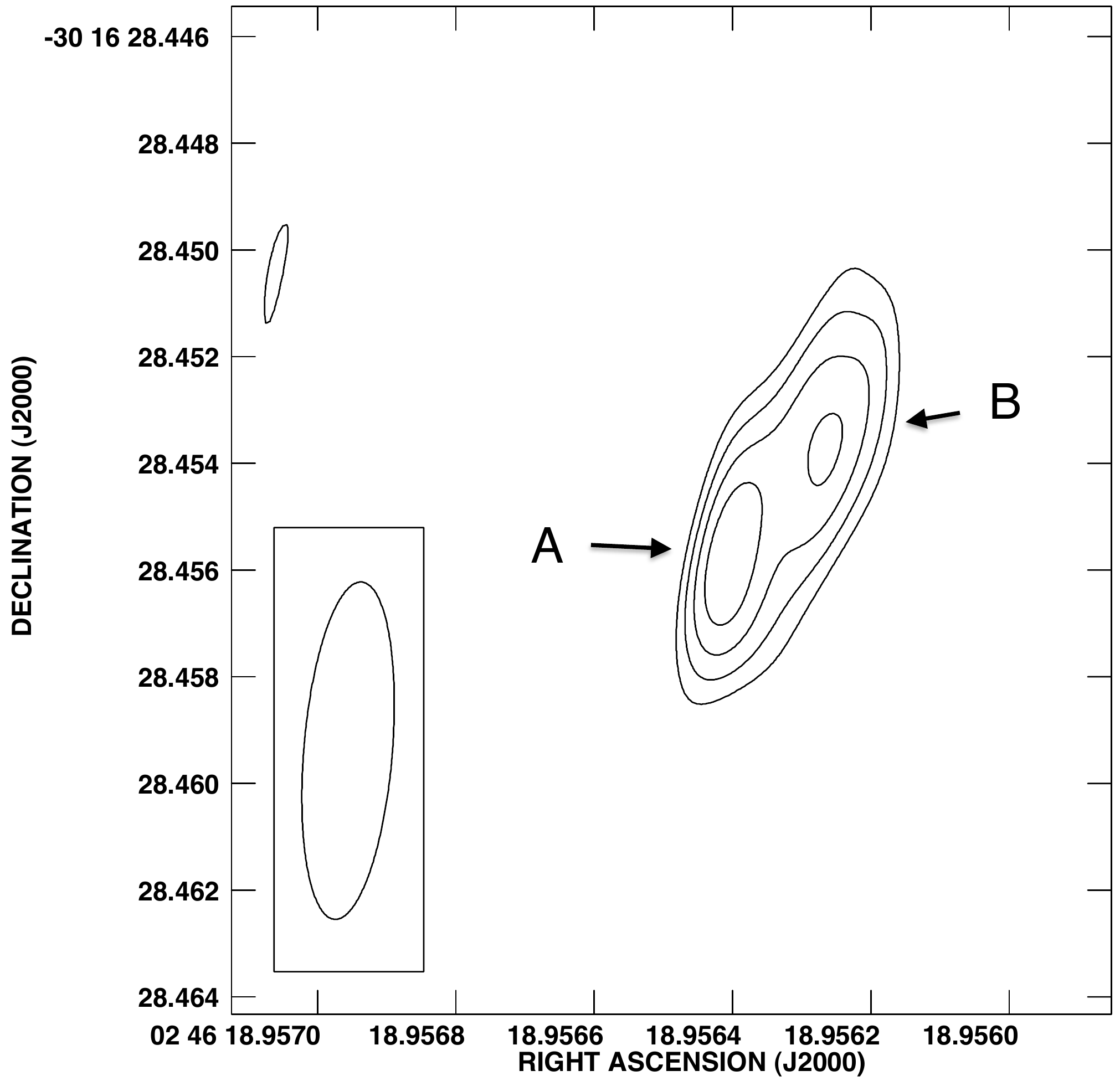}
 \includegraphics[scale=0.4]{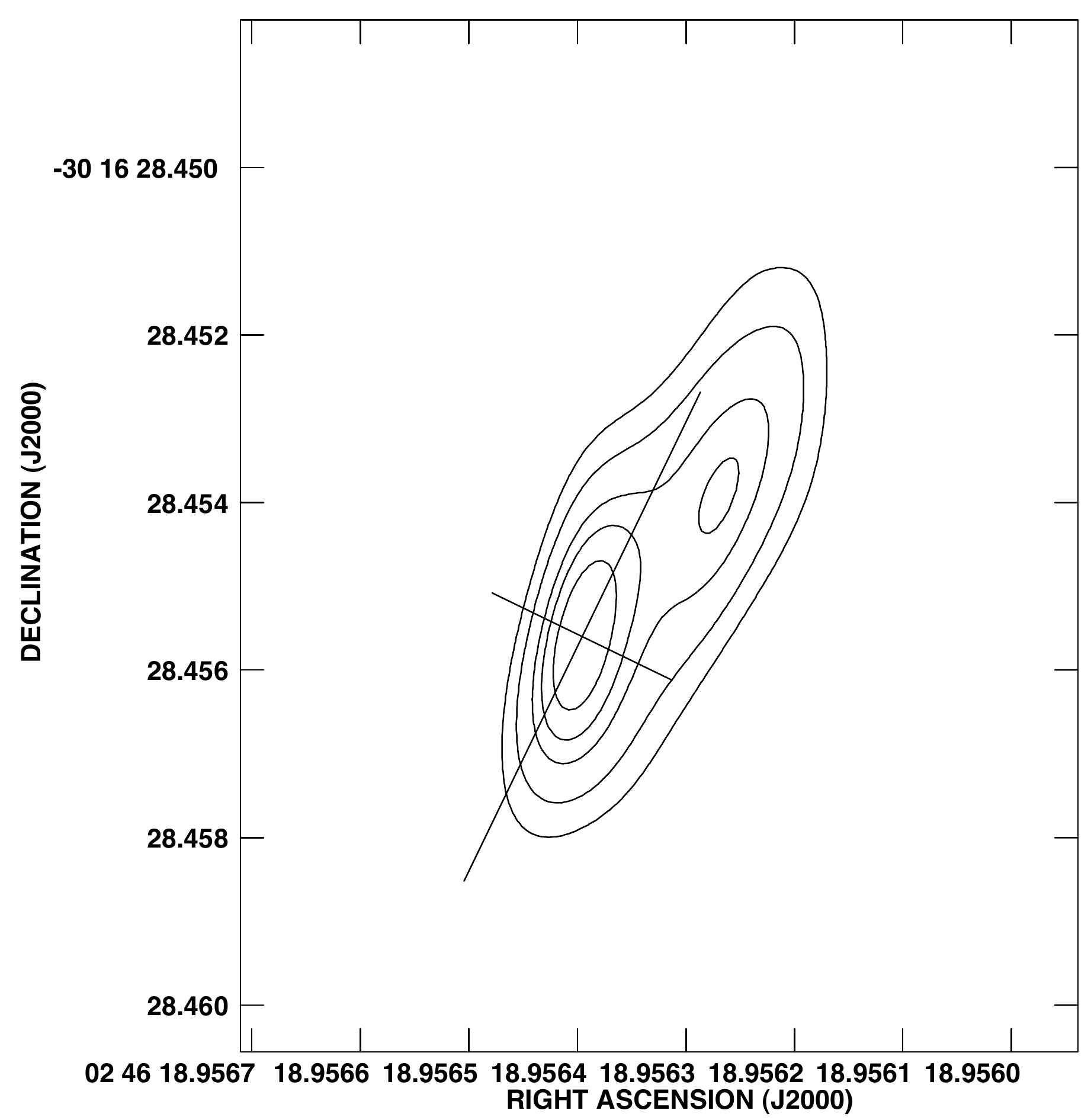} 
  \protect\caption[NGC1097]{VLBA image of NGC\,1097 at 8.4 GHz. The contours are (--3, 3, 4, 5, 6, 7) times the off-source rms noise of 0.13 mJy beam$^{-1}$. The restoring beam size is 6.34 mas $\times$ 1.66 mas, with the major axis of the beam oriented along a P.A. of $-4^{\circ}.6$. The peak flux density is 0.9 mJy beam$^{-1}$. The position and size of the core component derived from the two-dimensional Gaussian fitting is marked with a cross in the right panel in order to highlight the detection of extended emission. \label{ngc1097}}
\end{figure*}

\begin{figure*}
\includegraphics[scale=0.9]{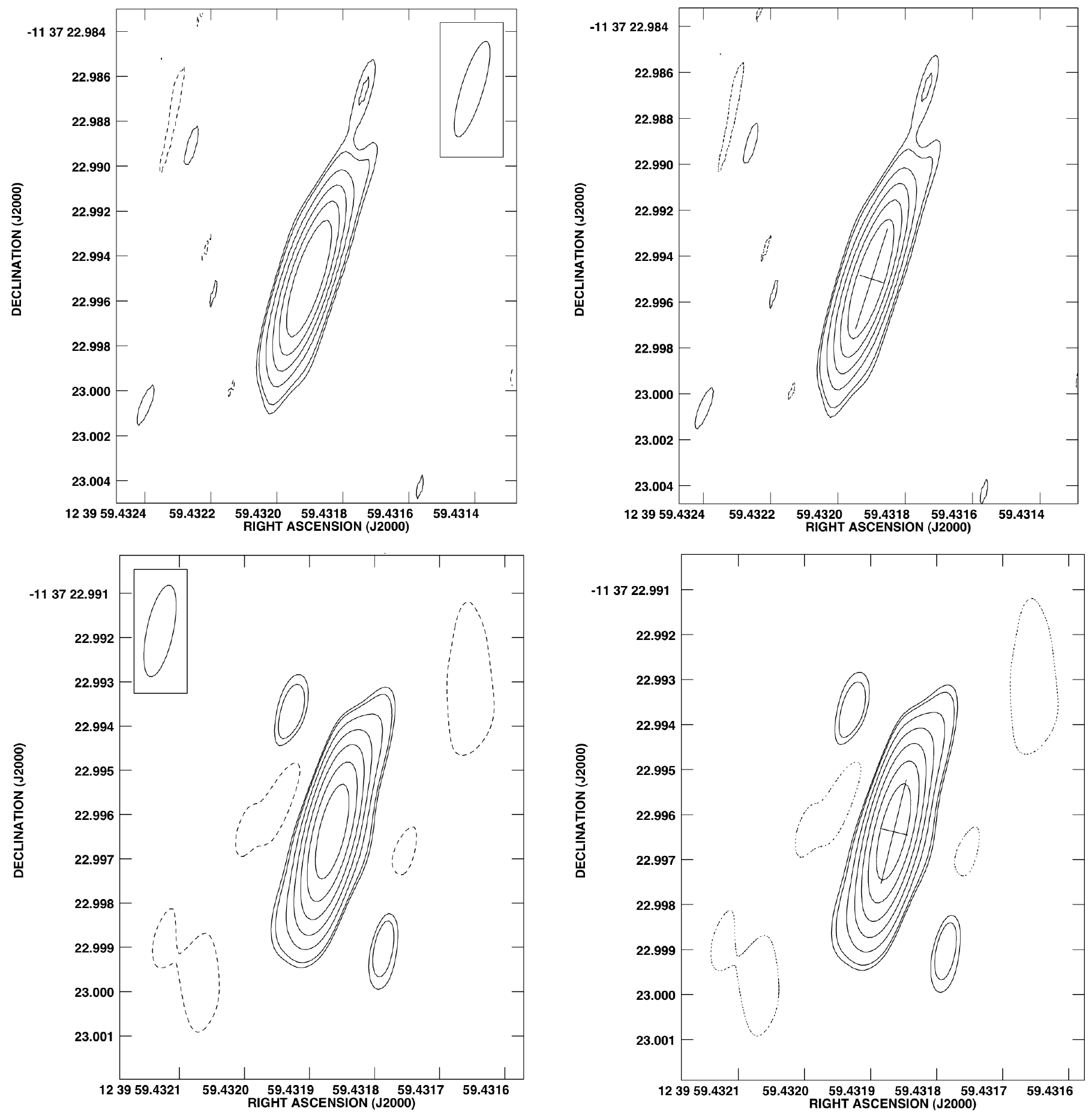}
 \protect\caption[NGC4594]{VLBA images of NGC\,4594 at 8.6 GHz (top) and 15.3 GHz (bottom). Contours are integer powers of $\sqrt{2}$ and start at three times the off-source rms noise of 0.4 mJy beam$^{-1}$ (at 8.6 GHz) and 0.3 mJy beam$^{-1}$ (at 15.3 GHz). The peak flux densities are 68.0 mJy beam$^{-1}$ and 73.5 mJy beam$^{-1}$, respectively. The beam size at 8.6 GHz is 4.43 mas $\times$ 0.97 mas oriented at a P.A. of $-16^{\circ}.9$. At 15.3 GHz, the beam has a size of 2.11 mas $\times$ 0.60 mas at a P.A. of $-12^{\circ}.1$. The position and size of the core component at each frequency derived from the two-dimensional Gaussian fitting is marked with a cross in the right panels in order to highlight the detection of extended emission. \label{ngc4594}}
\end{figure*}

\begin{figure}
\includegraphics[scale=0.35]{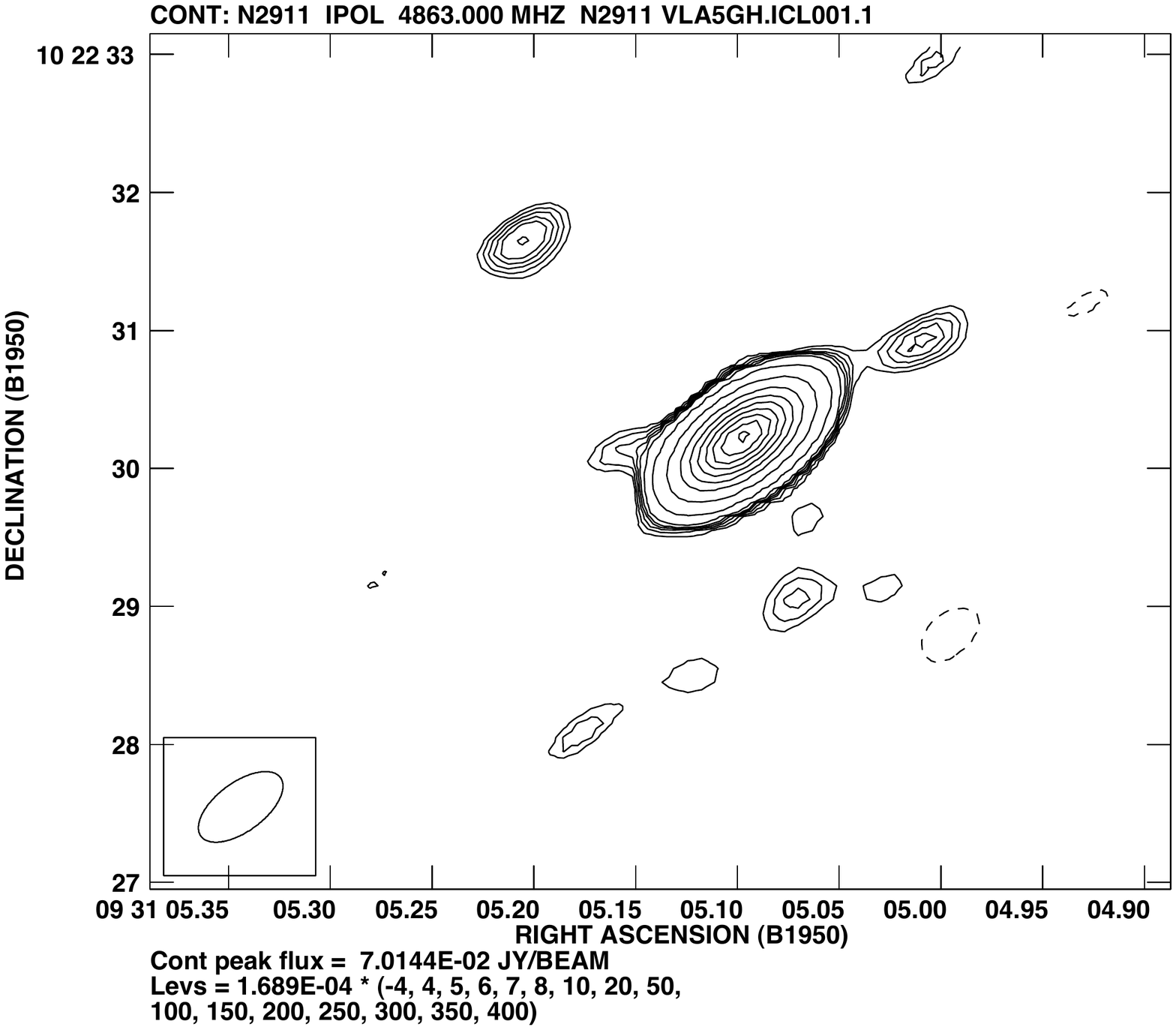}
 \includegraphics[scale=0.38]{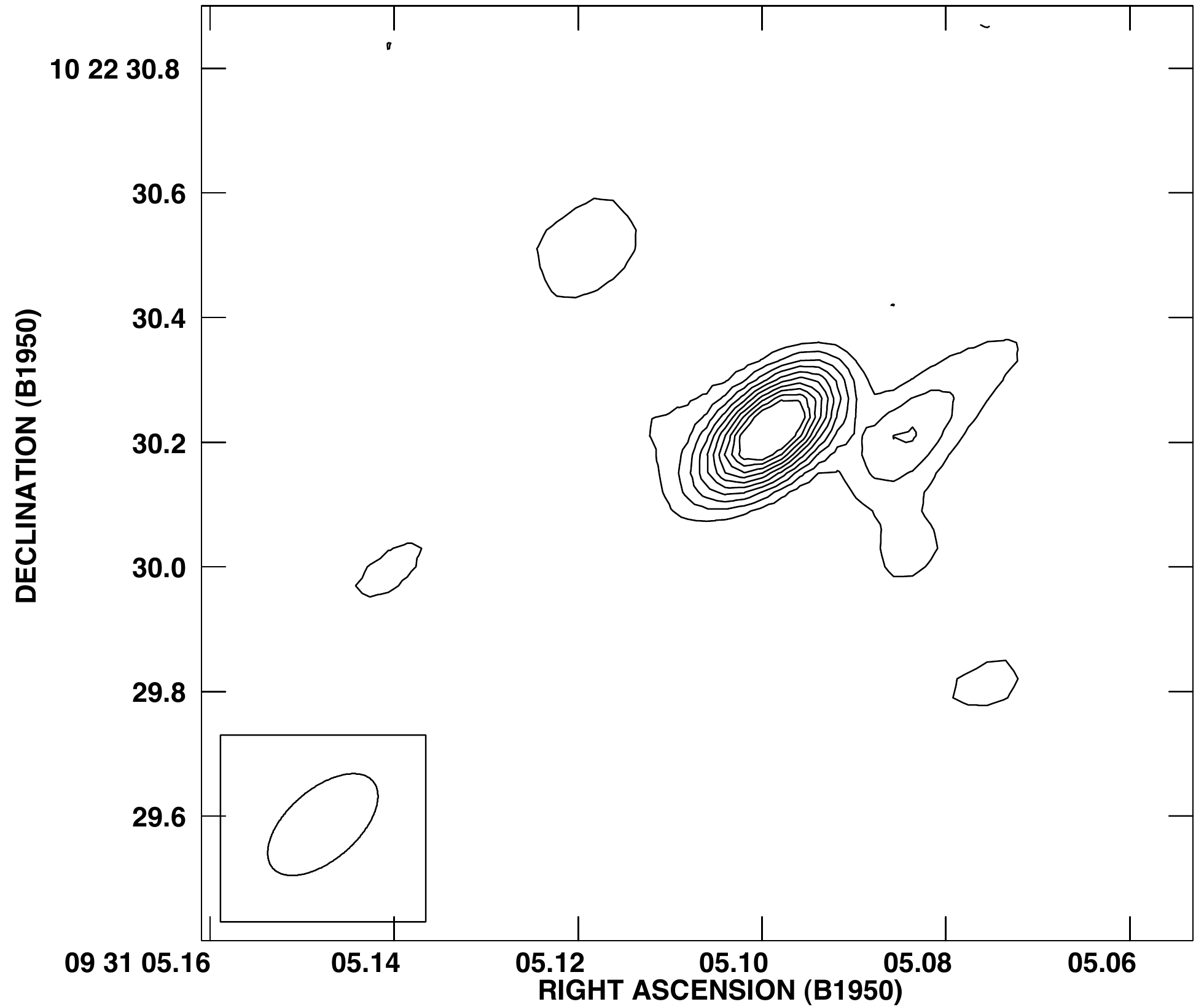}
 \protect\caption[NGC2911]{VLA-A array images of NGC\,2911 at 4.9 GHz (top) and 14.9 GHz (bottom). Contours are integer powers of $\sqrt{2}$ and start at three times the off-source rms noise of 0.2 mJy beam$^{-1}$ (at 4.9 GHz) and 0.6 mJy beam$^{-1}$ (at 14.9 GHz). The peak flux densities are 70.1 mJy beam$^{-1}$ and 21.5 mJy beam$^{-1}$, respectively. The beam size at 4.9 GHz is 0$\arcsec$.72 $\times$ 0$\arcsec$.35 oriented at a P.A. of $-53^{\circ}.6$. At 14.9 GHz, the beam has a size of 0$\arcsec$.21 $\times$ 0$\arcsec$.11 at a P.A. of $-49^{\circ}.0$.\label{ngc2911}}
\end{figure}

\begin{figure}
 \includegraphics[scale=0.38]{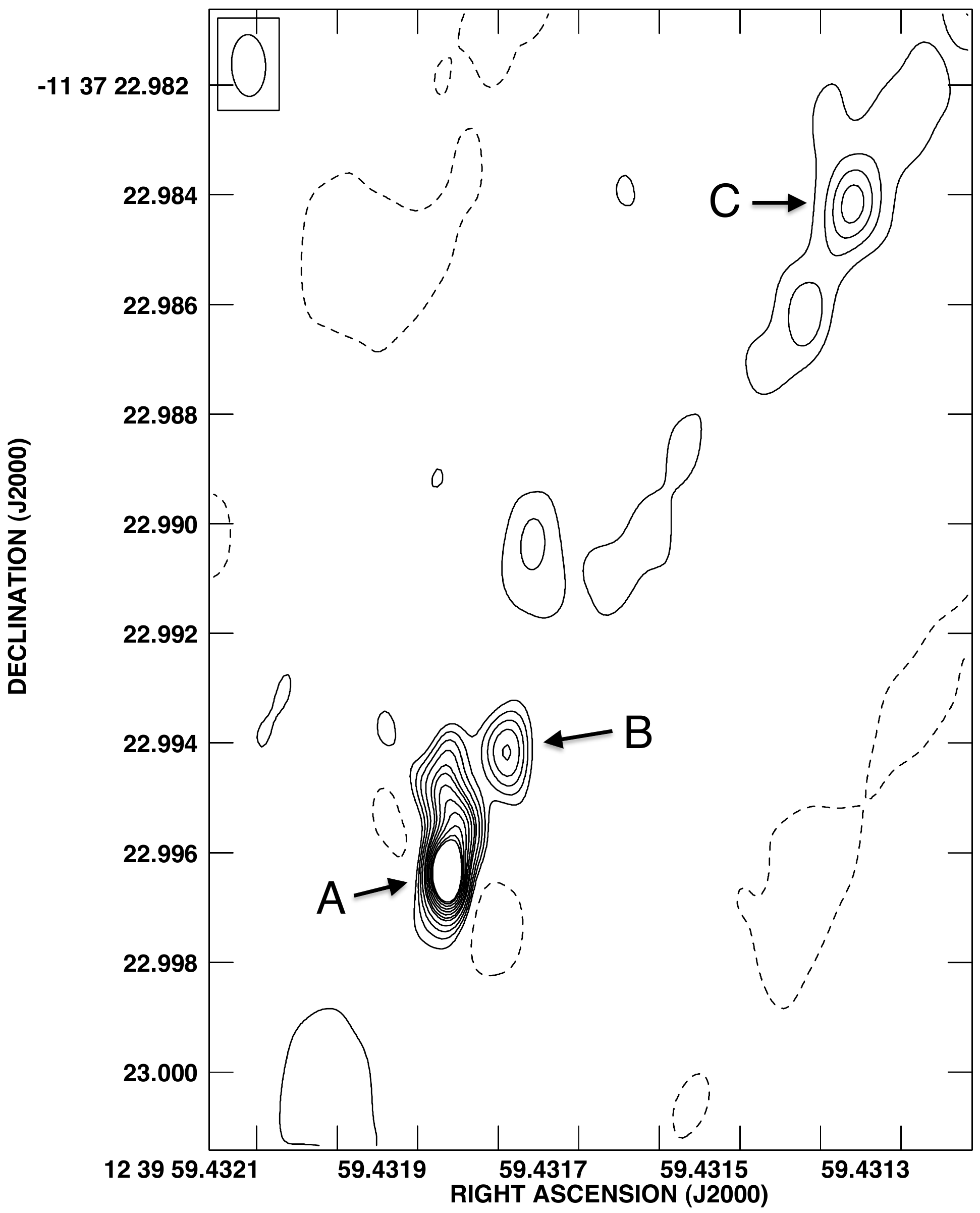}
  \protect\caption[NGC4594_24GHz]{VLBA image of NGC\,4594 at 23.8 GHz. Contours start at three times the off-source rms noise of 0.3 mJy beam$^{-1}$ and increase with factors of $\sqrt{2}$. The beam size is 1.13 $\times$ 0.62 oriented along a P.A. of $2^{\circ}.1$. The peak flux density is 8.9 mJy beam$^{-1}$. \label{ngc4594_23GHz}}
\end{figure}

Extended jet-like radio structures are detected above a signal-to-noise ratio (S/N)$\sim6$ for NGC\,1097, NGC\,2911, and NGC\,4594 (Figures~\ref{ngc1097}--\ref{ngc4594_23GHz}). The observed extended emission is consistent with the finding of integrated flux densities higher than peak densities and deconvolved sizes larger than beam sizes for all the studied sources and frequencies. The total angular and linear sizes, P.A., and flux density of the extended radio emission are given in Table~\ref{table2}. In those cases for which the extended emission is not clear enough from the radio maps (i.e., NGC\,1097, Figure~\ref{ngc1097}, left; NGC\,4594 at 8.6 and 15.3 GHz, Figure~\ref{ngc4594}, top left and bottom left, respectively), we plot the core component derived from the Gaussian model fitting in order to highlight the extended emission (Figure~\ref{ngc1097}, right for NGC\,1097; Figure~\ref{ngc4594}, top right and bottom right for NGC\,4594 at 8.6 and 15.3 GHz, respectively).

For NGC\,1097 two peaks of radio emission of S/N $\sim 7$ and S/N $\sim 6$, for the A and B components, respectively, are detected (Figure~\ref{ngc1097}), revealing the first extended jet-like structure of total linear size 0.7 pc and P.A.= $-23^{\circ}$. Assuming component A is associated with the central black hole (BH) and component B is associated with a radio lobe suggests the jet is partially oriented into our line of sight.  

For NGC\,2911 (Figure~\ref{ngc2911}) a bright core of S/N$\sim$400 and S/N$\sim$40 is detected at 4.9 and 14.9 GHz, respectively. In the 4.9 GHz image (Figure~\ref{ngc2911}, top), a component of S/N $\sim 9$ is observed northwest of the core emission, which is slightly resolved toward the east. The extended structure has a total linear size of $\sim$509 pc oriented along a P.A.= $-54^{\circ}$. An extended structure is also observed at 14.9 GHz (Figure~\ref{ngc2911}, bottom), though in this case the radio emission extends more to the west (P.A.= $-90^{\circ}$), where a component of S/N $\sim 9$ is detected. The total linear size of the 14.9 GHz extended structure is $\sim$81 pc.

For NGC\,4594, slightly resolved core radio emission is detected at 8.6 and 15.3 GHz (Figure~\ref{ngc4594}). At 23.8 GHz, we resolve for the first time the extended structure of a pc-scale jet: three distinct components of S/N $\geq 8$ are identified (labeled A, B, C in Figure~\ref{ngc4594_23GHz}) forming a linear structure of $\sim$6 pc oriented at P.A.= $-31^{\circ}$ (size and orientation measured from core A to component C). 
At the time of writing this paper, a multifrequency study of NGC\,4594 was published by \cite{2013ApJ...779....6H}, who report the same data at 15 GHz and 23 GHz as the ones reported here. The direction of our marginally extended core emission at 8.6 and 15.3 (Figure~\ref{ngc4594}) and of our extended structure at 23.8 GHz (Figure~\ref{ngc4594_23GHz}) is consistent with their results, although our 15.3 GHz and 23.8 GHz beam size and S/N is lower than in their analysis and no pc-scale jet is detected or reported by these authors at 23 GHz. These discrepancies can be explained by several factors: first, in this work we analyze the data obtained in the 2011 March 23 observations, while \cite{2013ApJ...779....6H} also use later observations; second, the images were obtained in our case by removing the longest baselines when imaging with AIPS and without applying self-calibration in order to improve the detection of extended emission, while in the self-calibration process performed with DIFMAP by \cite{2013ApJ...779....6H} all the baselines were included. 

The core flux measured at 4.9 and 14.9 GHz for NGC\,2911 allows us to derive a steep spectral index $\alpha = -1.1$. 
For NGC\,4594, an inverted spectral index $\alpha = 0.2$ is derived from the flux densities at 8.6 and 15.3 GHz. For NGC\,1097, the detection of extended radio emission in a single frequency does not allow the determination of the radio spectral index. However, an inverted spectral index $\alpha = 0.6$ is derived for this source from the core flux densities of 4 mJy (at 8.4 GHz) and 5.6 mJy (at 14.9 GHz) obtained by \cite{2010MNRAS.401.2599O} from VLA archival data. It should be noted that the spectral indices of NGC\,1097 and NGC\,4594 are derived from non-simultaneous observations and can be thus affected by variability effects (i.e., the use of non-simultaneous data can steepen the spectral index). This is indeed the case for NGC\,4594, for which \cite{2013ApJ...779....6H} find a flatter core spectral index ($\alpha = 0.14 \pm 0.02$) based on simultaneous VLBA observations at several frequencies. 

\section{Discussion}
\label{discussion}
A large 0$^{\prime\prime}$.15 resolution 15 GHz VLA survey of 162 LLAGN and AGN from the Palomar Sample was performed by Nagar et al. (\citeyear{2000ApJ...542..186N}, \citeyear{2002A&A...392...53N}, \citeyear{2005A&A...435..521N}) to search for compact radio emission in these sources. 
VLBA 5 GHz follow-up observations conducted with the aim of obtaining mas-resolution maps of all those LLAGN and AGN with VLA fluxes $>$2.7 mJy resulted in a 98\% detection rate of mas-scale radio emission (\citealt{2005A&A...435..521N}). The VLBA survey contained 42 LLAGN\footnote{We have excluded the sources NGC\,1275 and NGC\,4151 from the statistics as they are classified as AGN and not LLAGN.} (including Seyferts, LINERs, and transition objects), of which 18 (43\%) showed pc-scale ($\leq$100 pc) jets or larger (\citealt{2005A&A...435..521N}, table 2).
The LLAGN NGC\,3147, NGC\,3718, NGC\,4579, and NGC\,2911 remained compact in the 5 GHz VLBA observations; however, 5 GHz MERLIN\footnote{Multi-Element Radio-Linked Interferometer Network.} observations of NGC\,3147, NGC\,3718, and NGC\,4579 (\citealt{2007A&A...464..553K}) and the VLA observations of NGC\,2911 reported here are able to detect extended emission (see Section~\ref{detection}). The LLAGN NGC\,524, NGC\,3169, NGC\,3226, NGC\,3998, NGC\,4472, and NGC\,5866 also show slightly resolved radio cores in subarcsecond-scale radio maps consistent with (1) higher integrated flux density than peak flux density and (2) deconvolved size larger than the beam size (\citealt{2002A&A...385..425F}, 2004; \citealt{2004ApJ...603...42A}). We thus find that 67\% (28 out of 42) of the LLAGN of the Palomar Sample at subarcsecond resolution show pc-scale or larger radio jets. Among these sources with radio jets, 64\% correspond to LINERs (18 sources). 

Other LLAGN not included in the Palomar Sample (e.g., Arp 299-A, NGC\,1052, NGC\,1961, NGC\,4258, NGC\,7217) also show extended pc-scale radio emission (e.g., \citealt{2004A&A...420..467K}; \citealt{2007A&A...464..553K}; \citealt{2010A&A...519L...5P}; \citealt{2013ApJ...765...63D}), and in this work we report the first detection of extended pc-scale jets in the LINERs NGC\,1097 and NGC\,4594 (also not included in the Palomar Sample). All these sources cannot be added to the statistics as they do not belong to a flux-limited complete sample, but indicate that the number of LLAGN with pc-scale jets is probably much larger than the statistical values reported above. For instance, higher percentages of detection of small-scale extended jet emission are found in other samples like the NUGA survey (\citealt{2007A&A...464..553K}; 100\% jet detection rate) and, after including NGC\,1097 and NGC\,2911, eight out of the eight LLAGN of the southern sample of PR2010 also present pc-scale jet emission. 

The high detection rate of these mas-scale jets, together with the high brightness temperatures ($T_\mathrm{B} > 10^{6}$ K) radio cores and the flat to inverted spectral indices found for most of them (e.g., \citealt{2000ApJ...542..197F}; \citealt{2001ApJ...559L..87N}, \citeyear{2005A&A...435..521N}; \citealt{2007A&A...464..553K}; \citealt{2013ApJ...765...63D}; \citealt{2013ApJ...779....6H}), indicate that the central engine of these sources is non-thermal and associated with AGN. 
The finding that the subarcsecond-scale radio emission of these LLAGN is dominated by the jet is in agreement with multiwavelength high-spatial-resolution ($< 0.5$ arcsec) studies of LLAGN (e.g., \citealt{2012JPhCS.372a2006F}, in preparation) whose SEDs have been found to be well described by models dominated by synchrotron jet emission, thus confirming the necessity of a radio jet component when modeling the SED of these sources (e.g., \citealt{1999ApJ...516..672H}; \citealt{2001ApJ...562L.133U}; \citealt{2005MNRAS.360..119D}a; \citealt{2005ApJ...621..130W}; \citealt{2006ApJ...643..652N}; \citealt{2007ApJ...669...96W}).

\subsection{The Detection of Jet Radio Emission}
\label{detection}
The detection of extended jet emission in 67\% of the LLAGN of the Palomar Sample observed at subarcsecond resolution indicates that most LLAGN are energetic enough to power, at least, pc-scale jets (see also Section~\ref{jetpower}). The reason that their radio emission often appears compact is due to radio observations being not sensitive enough or of not high enough resolution.
This is the case, for instance, with NGC\,5354, whose core radio emission is unresolved in the 0$^{\prime\prime}$.2 resolution VLA map of \cite{2004A&A...418..429F} but shows double-sided extended emission in their higher-resolution VLBA observations. This extended emission is however not seen in shorter integration time VLBA maps.
Other clear examples are NGC\,3718 and NGC\,4579, which are not resolved in the 5 GHz VLBA survey of the Palomar Sample (with beam sizes  of 2.2 mas $\times$ 1.7 mas and 4.2 mas $\times$ 1.5 mas, respectively; \citealt{2005A&A...435..521N}) but do show extended jet emission in lower-resolution MERLIN observations (beam sizes 0$^{\prime\prime}.05 \times$ 0$^{\prime\prime}$.03 and 0$^{\prime\prime}.09 \times$ 0$^{\prime\prime}$.03, respectively) due to their higher S/N. Another example is NGC\,6500, which shows a compact core in the 5 GHz European VLBI Network (EVN) observations of \cite{2002A&A...385..425F}, but presents a linear extended structure in the higher resolution 5 GHz VLBA observations of \cite{2000ApJ...542..197F}. 

\begin{figure}[h!]
\includegraphics[scale=0.83]{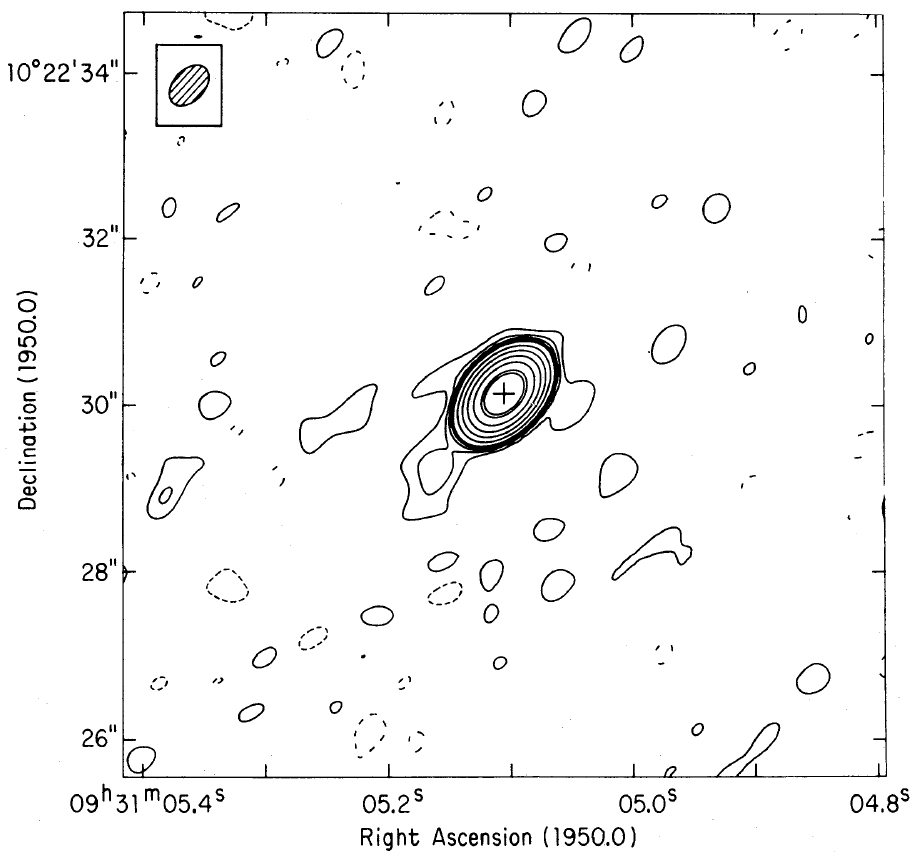}
 \protect\caption[NGC2911]{VLA-A array image of NGC\,2911 at 4.9 GHz from \cite{1984ApJ...287...41W}. The peak total intensity is 154.6 mJy beam$^{-1}$. The contours plotted are --3, --2, --1, 1, 2, 3, 4, 5, 10, 20, 40, 80, 160, and 200 times 0.25\% the peak intensity. The beam size is 0$\arcsec$.58 $\times$ 0$\arcsec$.38 oriented at a P.A. of $-43^{\circ}$. Figure and caption taken from \cite{1984ApJ...287...41W}. \label{wrobel}   }
\end{figure}

This is also evident in the LINERs NGC\,1097 and NGC\,2911 reported in this work. For NGC\,1097, unresolved core emission is detected with the VLA at 8.4 and 14.9 GHz at a resolution of 0$^{\prime\prime}.66 \times$ 0$^{\prime\prime}$.25 and 1$^{\prime\prime}.15 \times$ 0$^{\prime\prime}$.45, respectively (e.g., \citealt{2010MNRAS.401.2599O}). However, the core emission appears slightly extended in the 8.4 GHz radio map of \cite{2000MNRAS.314..573T} of resolution 0$^{\prime\prime}.59 \times$ 0$^{\prime\prime}$.25, while in the radio data presented in this paper, of mas-resolution, we detect two peaks of emission. 
For NGC\,2911, the core is slightly resolved in previous VLA campaigns (e.g., \citealt{1984ApJ...287...41W}; \citealt{1991AJ....101..362C}) but remains unresolved in VLBI observations due to lower S/N (e.g., \citealt{2002A&A...385..425F}). The core extends toward the northwest direction in the 4.9 GHz VLA map of \cite{1991AJ....101..362C}, which is consistent with the detection of a northwest component in our 4.9 GHz VLA map of slightly higher resolution (Figure~\ref{ngc2911}, top). A weak feature to the west of the core is also observed at 14.9 GHz (Figure~\ref{ngc2911}, bottom), which seems to be in agreement with a speculative feature that can be observed in the 4.9 GHz map of \cite{1984ApJ...287...41W} at a P.A. of $-90^{\circ}$ (see Figure~\ref{wrobel}). The eastern component detected at 14.9 GHz (Figure~\ref{ngc2911}, bottom) is however not detected by \cite{1984ApJ...287...41W}, who detect only a compact component at 15 GHz, possibly due to a higher rms noise.
In Section~\ref{results} we reported a spectral index of $\alpha$=1.1 for the slightly resolved core emission of NGC\,2911, which is much steeper than the flat $\alpha$=0.21 obtained by \cite{2002A&A...385..425F} from 5 GHz EVN observations of higher angular resolution than the VLA observations reported here. This, together with the size of a few hundreds of parsec (Table~\ref{table2}) of the extended radio emission that we detect at 4.9 GHz, indicates that with the VLA we are indeed detecting extended jet emission and not only the core of NGC\,2911.

\subsection{Large-scale Jet Emission and Jet Alignment}
\label{kpcscale}
Only $\sim$1/3 of the LLAGN of the Palomar and PR2010 samples with pc-scale emission also present larger-scale (100 pc to kpc) extended jets. This is the case of Cen~A (e.g., \citealt{1992ApJ...395..444C}), M87 (e.g., \citealt{2005AJ....130.1389L}), NGC\,1052 (e.g., \citealt{1984ApJ...284..531W}; \citealt{2004A&A...420..467K}), NGC\,3079 (e.g., \citealt{1978A&A....63..199S}; \citealt{1983ApJ...267L...5H}; \citealt{1995ApJ...446..602B}), NGC\,3169 (e.g., \citealt{1987A&AS...70..517H}), NGC\,3998 (e.g., \citealt{1980A&AS...41..151H}; \citealt{1984ApJ...287...41W}), NGC\,6500 (\citealt{1989A&A...208...14U}), NGC\,315, NGC\,1275, NGC\,4261, NGC\,4374, and NGC\,7626 (e.g., \citealt{2005A&A...435..521N}), and NGC\,4594 (\citealt{2006AJ....132..546G}). 

For Cen~A, M87, NGC\,1052, NGC\,1275, NGC\,315, NGC\,3169, NGC\,4261, NGC\,4374, and NGC\,7626 the radio jet is collimated from sub-pc to kpc scales (i.e., the P.A. of the pc-scale jet is consistent with that of the kpc-scale radio structure), while for NGC\,3079 and NGC\,6500 the mas-scale jet is clearly misaligned with the two-sided extended arcsec emission (e.g., \citealt{2000PASJ...52..421S}; \citealt{2002A&A...385..425F}). Such jet misalignments can be explained either by intrinsic variations in the jet orientation (e.g., caused by jet precession or a BH merger; \citealt{2002Sci...297.1310M}; \citealt{2004ApJ...616L..99C}; \citealt{2011A&A...527A..38M}, \citeyear{2012A&A...544A..36M}) or by jet interaction with the interstellar medium (ISM) from pc to larger scales (e.g., \citealt{1996ApJ...458..136G}, \citeyear{1996ApJ...462..740G}; \citealt{2001ApJ...552..144D}; \citealt{2004A&A...417..925M}). The latter scenario is favored for NGC\,6500 given that its pc-scale radio jet is aligned with the extended ionized gas emission (\citealt{1996MNRAS.281.1105G}; \citealt{2002A&A...385..425F}).
If interactions with the ISM affect the jet orientation, from pressure gradient arguments jets should on average bend toward the minor axis of the host galaxy (\citealt{2006AJ....132..546G}; see also review by \citealt{2012RAA....12..127G}). This is the case with NGC\,3079, whose two 15 kpc radio lobes extend along the minor axis of the galaxy. However, this argument is strongly affected by projection effects (see \citealt{2006AJ....132..546G} for further discussion).

For NGC\,4594, a connection between the pc-scale and the kpc-scale jet emission has not been mentioned in previous studies of its mas-scale jet (e.g., \citealt{2013ApJ...779....6H}). NGC\,4594 presents a bright ($\sim$100 mJy; e.g., \citealt{1984A&A...134..207H}; \citealt{1988A&A...202...35B}) radio continuum that extends along the major axis of the galaxy. A fainter kpc-scale (3.8 kpc) linear radio structure oriented northwest of the galaxy minor axis (P.A. = $-38^{\circ}$) was detected by \cite{2006AJ....132..546G}, who suggested that it originates from an AGN jet.  
The orientation of the 23.8 GHz pc-scale jet reported here (P.A. = $-31^{\circ}$) is consistent with the kpc-scale radio structure from \cite{2006AJ....132..546G}, indicating that the nucleus of this source is able to collimate a jet from pc to kpc scales as in M87 or NGC\,1052. On the other hand, the pc-scale jet of NGC\,4594 is misaligned with the direction of the ionized gas emission, which is elongated along the major axis of the host galaxy (\citealt{Prieto2014}). The jet might thus not be so strongly affected by ISM interactions, unlike NGC\,6500, which is consistent with the alignment of the small- to the large-scale radio jet of NGC\,4594.

The large-scale jet emission is thus aligned to the pc-scale jet in 10 out of 13 (77\%) LLAGN for which both pc- and kpc- scale jets are detected. Although the small number of sources does not allow a significant statistical study, it is interesting to note that a similar percentage of jet alignments ($\sim$70\%, 8 out of 12 sources) was found in a survey of kpc-scale radio outflows in radio-quiet AGN (\citealt{2006AJ....132..546G}). 


\subsection{Jet Power}
\label{jetpower}
The non-detection of large-scale jets in most LLAGN suggests that their pc-scale jet is frustated in a region of a few parsecs either because it is
uncollimated or due to interaction with the ambient medium, which would also explain the typically curved or bent radio morphology (e.g., \citealt{2005A&A...435..521N}). This could be the case with NGC\,1097, for which kinematical studies of the nuclear H$_{2}$ molecular gas reveal an increase of the dispersion at tenths of parsec scales that could result from the interaction of the gas with the radio jet (\citealt{2013ApJ...763L...1M}).

To investigate this further, we compile in Table~\ref{powerLbol} the jet mechanical luminosity or kinetic power ($Q_\mathrm{jet}$) and the bolometric radiative luminosity ($L_\mathrm{bol}$) of all the sources in the southern sample of bright nearby LLAGN of PR2010. The BH mass ($M_\mathrm{BH}$), from which the Eddington luminosity ($L_\mathrm{Edd}$) is derived, the Eddington-scaled accretion rate (defined as $\lambda=(Q_\mathrm{jet}+L_\mathrm{bol})/L_\mathrm{Edd}$), the Eddington ratio ($l_\mathrm{Edd}=L_\mathrm{bol}/L_\mathrm{Edd}$), and the largest observed jet size are also reported in Table~\ref{powerLbol}.

Studies of the jet power and bolometric luminosity of the LLAGN in the Palomar Sample indicate that these sources have a highly sub-Eddington $l_\mathrm{Edd}$ and that their accretion is dominated by the jet mechanical power (e.g., \citealt{2005A&A...435..521N}). In a larger sample of more than 10,000 radio sources, \cite{2012MNRAS.421.1569B} find that LLAGN have accretion rates $\sim10^{-3}$ Eddington, consistent with LLAGN being radiatively inefficient and fueled by gas from the hot X-ray halo of the host galaxy. These statistical studies are based on crude estimates of $L_\mathrm{bol}$ (i.e. from its correlation with the X-ray or [OIII] luminosity; e.g., \citealt{1999ApJ...516..672H}; \citealt{2001ApJ...558..561U}; \citealt{2004ApJ...613..109H}) and of $Q_\mathrm{jet}$ (i.e. assuming an average inclination in the case of the Palomar Sample; \citealt{2005A&A...435..521N}). A detailed study of individual objects is required to test the statistical results. This can be performed for the LLAGN in the sample of PR2010, for which the $L_\mathrm{bol}$ has been derived from the high-spatial-resolution core SED (e.g., \citealt{2010MNRAS.402..724P}; \citealt{2012JPhCS.372a2006F}, private communication) and the $Q_\mathrm{jet}$ is directly estimated, for some of them (Cen A, M87, and NGC\,1052), from the observed bubbles in X-ray emitting gas or from maser emission. The values of $L_\mathrm{bol}$ and $Q_\mathrm{jet}$ of the sources in the PR2010 sample are thus the most reliable that can be achieved, constituting a much better estimate than those values inferred from bolometric corrections, as these have been proven to be wrong by a huge factor (e.g., \citealt{2010MNRAS.402..724P}; \citealt{2014MNRAS.438..672N}). For those cases where the $Q_\mathrm{jet}$ could not be derived from X-ray cavities, we use the correlation between the core radio luminosity at 5 GHz and the $Q_\mathrm{jet}$ of \citeauthor{2007MNRAS.381..589M} (2007, Equation.~(6)), which  has the minimum scatter (0.4 dex) and has been corrected for relativistic beaming. For NGC\,1097, the core radio luminosity at 5 GHz is derived from the 8.4 GHz radio flux densities of component A\footnote{It is not clear whether component A or B is the core  of NGC\,1097 as the two peaks of emission in the core structure had never been detected before and no spectral information for them is available. We assume that A is the core position when calculating the jet power. Using the flux density of component B to derive the core radio luminosity, a value of $Q_\mathrm{jet} = 2.0 \times 10^{41}$ erg s$^{-1}$ would be obtained.} properly scaled using $\alpha = 0.6$ (see Section~\ref{results}). For NGC\,4594, we use the 5 GHz core flux density reported by \cite{2013ApJ...779....6H}.

\begin{table*}
\centering
\begin{threeparttable}
\caption{Properties of the PR2010 Sample \label{powerLbol}}
\scriptsize
\begin{tabular}{lccccccccc}
\hline
\hline 
 LLAGN	& $D_\mathrm{L}$   & Type &   $Q_\mathrm{jet}$  &  $L_\mathrm{bol}$  &  	$M_\mathrm{BH}$	&	$\lambda$		&$l_\mathrm{Edd}$		&  Largest Jet Size  & Ref.      \\
               &     (Mpc)              &       & (erg s$^{-1}$)	    &   (erg s$^{-1}$)	     &	          (M$_{\odot}$)	       &     				& 					&	(kpc)            &	        \\
 (1)		& 	(2)			& (3)	&	(4)				&   (5)			&    (6)				&	(7)			&	(8)				&    (9)		&  (10)  \\
\hline
NGC\,1097 & 	 18	 & L1    & $1.5 \times 10^{41}$    &	$3.8 \times 10^{41}$&	$1.3 \times 10^{8}$	& $3.3 \times 10^{-5}$	& $2.4 \times 10^{-5}$	& 	0.1	&    1,2,3,4	\\
NGC\,2911 & 	 40	 & L2    & $5.9 \times 10^{42}$    &	$2.0 \times 10^{41}$&	$1.9 \times 10^{8}$	& $2.5 \times 10^{-4}$	& $8.4 \times 10^{-6}$	&	0.5  		&    1,5,6		 \\
NGC\,4594 &	  9	 & L2    & $2.1 \times 10^{42}$ 	&	$1.9 \times 10^{41}$&	$1.0 \times 10^{9}$	& $1.9 \times 10^{-5}$	& $1.5 \times 10^{-6}$	&	3.8		&    1,6,7,8        \\
Cen A	  & 	3.4	 & Sy2  & $6 \times 10^{42}$ 	&	$2.8 \times 10^{42}$&	$4.5 \times 10^{7}$	& $1.6 \times 10^{-3}$	& $5.0 \times 10^{-4}$	&	2		&    2,9,10,11	   \\
M87		  & 	17	 & L1   & $8 \times 10^{42}$  	&	$1.2 \times 10^{42}$&	$6.6 \times 10^{9}$	& $1.1 \times 10^{-5}$	& $1.5 \times 10^{-6}$	&	1.5	&    6,12,13,14 	   \\
NGC\,1052 & 	18	 & L1.9 & $6 \times 10^{42}$	&	$5.0 \times 10^{42}$&	$1.0 \times 10^{8}$	& $8.8 \times 10^{-4}$	& $4.0 \times 10^{-4}$	&	3		&    5,6,15,16   	 \\
NGC\,1386 & 	15.3	 & Sy2  & $1.2 \times 10^{42}$	&	$2.9 \times 10^{42}$&	$1.4 \times 10^{8}$	& $2.3 \times 10^{-4}$	& $1.7 \times 10^{-4}$	&	0.1 		&    6,17,4  	 \\
NGC\,3169 & 	24.7	 & L2    & $1.5 \times 10^{42}$	&	$7.4 \times 10^{41}$&	$9.8 \times 10^{7}$	& $1.8 \times 10^{-4}$	& $6.0 \times 10^{-5}$	&	0.5		&    5,6,18   	 \\
\tableline
\end{tabular}
\begin{tablenotes}
      \small
      \item \textbf{Notes.}~(1) Source name; (2) luminosity distance; (3) LLAGN type; (4) jet kinetic power; (5) bolometric luminosity; (6) black hole mass; (7) Eddington-scaled accretion rate ($\lambda=(Q_\mathrm{jet}+L_\mathrm{bol})/L_\mathrm{Edd}$); (8) Eddington ratio ($l_\mathrm{Edd}=L_\mathrm{bol}/L_\mathrm{Edd}$); (9) largest projected jet size; (10) references. The LLAGN types stand for L: LINER, Sy: Seyfert. REFERENCES:~(1) This work; (2) \cite{2010MNRAS.402..724P}; (3) \cite{2009ApJ...702..114D}; (4) \cite{2000MNRAS.314..573T}; (5) \cite{2013ApJ...763L...1M}; (6) \cite{2012JPhCS.372a2006F}, private communication; (7) \cite{2013ApJ...779....6H}; (8) \cite{1996ApJ...473L..91K}; (9) \cite{2009ApJ...698.2036K}; (10) \cite{2007ApJ...671.1329N}; (11) \cite{1983ApJ...273..128B}; (12) \cite{2013MNRAS.432..530R}; (13) \cite{2011ApJ...729..119G}; (14) \cite{1989ApJ...342..128B}; (15) \cite{2005ApJ...620..145K}; (16) \cite{1984ApJ...284..531W}; (17) \cite{2003A&A...400...41S}; (18) \cite{1987A&AS...70..517H}.
    \end{tablenotes}
\end{threeparttable}
\end{table*}

The values compiled in Table~\ref{powerLbol} show that the $Q_\mathrm{jet}$ of NGC\,2911 and NGC\,4594 are comparable to that of Cen~A, M87, and NGC\,1052, while NGC\,1097 presents a more than one order of magnitude lower $Q_\mathrm{jet}$ than that of the powerful radio galaxies (i.e. those with $Q_\mathrm{jet} > 6 \times 10^{42}$ erg s$^{-1}$, Cen A, M87, NGC\,1052). According to this comparison, NGC\,2911 and NGC\,4594 are thus powerful enough to collimate their radio jets to large scales as in Cen~A, NGC\,1052 or M87, which is indeed consistent with the detection of a $\sim$500 pc jet in NGC\,2911 (Figure~\ref{ngc2911}) and of a 3.8 kpc jet in NGC\,4594 (\citealt{2006AJ....132..546G}). The jet of NGC\,1097, on the contrary, does not reach 100 pc\footnote{the largest detected jet size for NGC\,1097 is $\sim$90 pc, measured on the radio map of \cite{2000MNRAS.314..573T}.}, which is in agreement with its low value of jet power ($Q_\mathrm{jet} < 10^{42}$ erg s$^{-1}$). This is also observed for the second LLAGN in Table~\ref{powerLbol} with the lowest jet power (NGC\,1386, $Q_\mathrm{jet}\sim10^{42}$ erg s$^{-1}$), which does also not show a large-scale ($>$100 pc) radio jet.

It should be noted that the jet sizes quoted in Table~\ref{powerLbol} correspond to projected sizes. This does not affect our results, as most sources are either of type 2, viewed nearly edge-on, and/or their kpc-scale jets extend to both sides of the nuclear region at the largest kpc scales (i.e. NGC\,4594, Cen~A, NGC~1052, M87, and NGC\,3169). None of these is the case for NGC\,1097, which is a type 1 LLAGN whose host galaxy is closer to a face-on view. However, the physical jet size of NGC\,1097 is $\sim$100 pc even when considering an inclination angle of $45^{\circ}$. The results obtained are thus not affected by projection effects.

Those LLAGN in Table~\ref{powerLbol} with large-scale jet radio emission (NGC\,2911, NGC\,4594, Cen~A, M87, NGC\,1052, and NGC\,3169) have values of $Q_\mathrm{jet}$ more than two times larger than their $L_\mathrm{bol}$, while for those LLAGN with small-scale radio jets ($\leq$100 pc; NGC\,1097 and NGC\,1386) $Q_\mathrm{jet}$ is smaller than $L_\mathrm{bol}$ but comparable to it when considering the 0.4 dex scatter in the estimate of $Q_\mathrm{jet}$ (\citealt{2007MNRAS.381..589M}). Accounting for uncertainties of a factor up to two, $Q_\mathrm{jet}$ is still comparable to $L_\mathrm{bol}$ for all sources regardless of the presence of a large-scale or pc-scale jet. Therefore the jet kinetic energy contributes at least as much as the radiative energy in these sources. It should be noted that the $Q_\mathrm{jet}$ estimated from the work done by the jet in opening the X-ray cavities is a lower limit to the total jet power, since part of the jet energy must be used to compress the X-ray gas as the jet advances in the ISM (e.g., \citealt{2012A&A...545L...3C}; \citealt{2012A&A...544A..56L}). The nuclear energetics of the LLAGN in the PR2010 sample are thus mostly dominated by the jet power, which is in agreement with the statistical results for the LLAGN in the Palomar Sample (\citealt{2005A&A...435..521N}). 

We find $l_\mathrm{Edd}$ ranging $10^{-6}-10^{-4}$ for the sources in the PR2010, which is consistent with the results obtained by \cite{2005A&A...435..521N} for the LLAGN in the Palomar Sample. The low Eddington ratios typically found in LLAGN (e.g., see also \citealt{2008ARA&A..46..475H}; \citealt{2010ApJS..187..135E}; \citealt{2011A&A...527A..23M}) can be owing to a very low supply of cold gas to the accretion radius. In those LLAGN with radio jets, this 
could be explained by the finding that, in most cases, the jet radio emission is confined to pc scales: the deposition of most of the jet power within the central parsecs can reduce the inflow of matter around the accretion radius and thereby decrease the accretion rate (e.g., \citealt{2001ApJ...547..731D}; \citealt{2005A&A...435..521N}). This can be even more significant in those sources with kpc-scale jets, for which the injection of energy into the galactic ISM can in addition hamper the arrival of any cooling flow to the surroundings of the accretion disk (e.g., \citealt{2001ApJ...547..731D}; \citealt{2006ApJS..166....1H}).    

For most sources in Table~\ref{powerLbol} (those for which $Q_\mathrm{jet}$ is more than two times larger than $L_\mathrm{bol}$) there is an increase in the Eddington ratio of nearly one order of magnitude when including the jet kinetic power to the total (kinetic plus radiated; $Q_\mathrm{jet}$ + $L_\mathrm{bol}$) emitted luminosity. The Eddington-scaled accretion rate is however still highly sub-Eddington ($\lambda < 10^{-3}$) for all sources, suggesting that, apart from being radiatively inefficient, these LLAGN are either inefficient accretors (i.e., no matter arrives at the accretion radius) or efficient advectors (i.e., they efficiently advect the matter that arrives to the accretion radius). The average Eddington-scaled accretion rate is $\lambda < 10^{-4}$, which is lower than the peak of $10^{-3}$ in the distribution of $\lambda$ found by \cite{2012MNRAS.421.1569B} for a large sample of $>$10,000 LLAGN. This difference in $\lambda$ can be explained by the following: (1) our estimates of $L_\mathrm{bol}$ from the high-resolution core SEDs are more reliable than the $L_\mathrm{bol}$ estimated by \cite{2012MNRAS.421.1569B} using a bolometric correction factor, which can yield an overestimated or wrong $L_\mathrm{bol}$ (the $L_\mathrm{bol}$ estimated from high-resolution SEDs is several orders of magnitude smaller than those estimated from low-resolution data; e.g., \citealt{2010MNRAS.402..724P}; see also \citealt{2014MNRAS.438..672N}); (2) because of the former, $\lambda$ can deviate from lower values if the statistical samples include too many objects with overestimated $L_\mathrm{bol}$.

Finally, we note that the increase in the Eddington ratio when including the jet kinetic energy should be even larger for those sources whose $Q_\mathrm{jet}$ has been estimated from X-ray cavities, since in these cases the values of $Q_\mathrm{jet}$ constitute lower limits to the jet power. Attempts to estimate the total jet power are model dependent. We are attempting to estimate it from the modeling of the SED of these sources, which appears very much consistent with a pure jet emission (Fernandez-Ontiveros et al. in preparation).

\section{Conclusions}
\label{conclusions}
LLAGN were suggested to be typically associated with unresolved radio sources even at mas scales; however, we find that 67\% of the LLAGN with subarcsecond radio emission of the Palomar Sample present extended pc-scale jets or slightly resolved core emission when observed with sufficient angular resolution and sensitivity. Most LLAGN are thus powerful enough to eject radio jets at least at pc scales. We also report the detection of extended jets in the LLAGN/LINERs NGC\,1097, NGC\,2911, and NGC\,4594 based on the analysis of archival VLA and VLBA data. For NGC\,4594, the pc-scale jet is resolved, for the first time, into several components. With these detections, all the sources (eight out of eight) in the sample of the brightest nearby LLAGN in the southern hemisphere of PR2010 present parsec-scale or larger radio jets. 

The extended detected structures of the LLAGN in the Palomar and the PR2010 samples, together with the derived high brightness temperatures and radio spectral indices, indicate a clear association of the radio emission with AGN. This reinforces the long debated scenario in which LLAGN are also powered by accretion onto a massive BH as AGN but have a different accretion mode (i.e., are radiatively inefficient), and supports those models in which the subarcsecond SED of these sources is dominated by synchrotron jet emission. The pc-scale jet emission is found to be aligned with the large (100 pc to kpc) jet emission in 10 out of 13 sources for which both a small- and large-scale jet is detected, indicating that the nuclear engine of these sources is powerful enough to collimate the jets from pc to kpc scales. The misalignment between the small and large-scale jet emission for the other sources can be explained by jet interactions with the ISM.

The high-spatial-resolution core SED available for the LLAGN in the PR2010 sample together with their jet power, either directly measured from X-ray cavities or derived from their core radio luminosity, allows us to derive the Eddington-scaled accretion rate and Eddington ratio of these sources with minimum ad-hoc assumptions. For the eight sources in the PR2010 sample we obtain that:
\begin{itemize}
\item A general trend is observed between jet power and jet size: those powerful radio galaxies with the largest radio jets (Cen A, M87, NGC\,1052) are also the ones with largest values of jet power, while the lowest values correspond to those LLAGN whose largest observed jet size is $\leq$ 100 pc. All the sources with jet power $>$ 10$^{42}$ erg s$^{-1}$ do show large-scale ($>$ 100 pc) radio jets.

\item The jet power is comparable to or larger than the bolometric luminosity for all LLAGN, in agreement with statistical studies of larger LLAGN samples. Our study of individual sources reveals the presence of radio jets when these sources are observed with enough resolution and sensitivity, indicating that the jet kinetic power contributes at least as much as the radiative output regardless of the jet size. The nuclear energetics of LLAGN is thus dominated by the jet kinematic output. 

\item The Eddington ratio ranges $10^{-6}$-$10^{-4}$ and is thus highly sub-Eddington. When including the jet power to the total emitted luminosity (radiated plus kinetic), the Eddington-scaled accretion rate increases nearly one order of magnitude but is still $<10^{-4}$ for nearly all sources. These results indicate that LLAGN are not only unequivocally inefficient radiators, the main channel for releasing their energy being the jet, but also either inefficient accretors or very efficient advectors. Even with the current best estimate of kinetic power and bolometric luminosity, they remain highly sub-Eddington.
 \end{itemize}

\section{Acknowledgements}
The authors are grateful for the insightful suggestions of the anonymous referee. 
The authors are very thankful to J.A. Fern\'andez Ontiveros for providing the bolometric luminosities. M.M. acknowledges financial support from AYA2011-25527. M.A.P. acknowledges Martin Krause and Klaus Dolag for motivating discussions.

\bibliographystyle{mn2e} 
\bibliography{~/Documents/referencesALL}

\end{document}